\documentclass[seceq]{ptptex}

\usepackage{graphicx}

\newcommand{\bfk}{\mbox{{\boldmath $k$}}}

\newcommand{\bfgamma}{\mbox{{\boldmath $\gamma$}}}

\title{
Pseudogap of Color Superconductivity
}

\author{
Masakiyo \textsc{Kitazawa}$^a$,
Tomoi \textsc{Koide}$^b$,
Teiji \textsc{Kunihiro}$^c$ and
Yukio \textsc{Nemoto}$^d$
}

\inst{
$^a$ Deparment of Physics, Kyoto University, Kyoto, 606-8502, Japan,\\
$^b$ Institute f\"ur Theoretische Physik, J.W.Goethe Universit\"at, D-60054
Frankfurt, Germany,\\
$^c$ Yukawa Institute for Theoretical Physics, Kyoto University, Kyoto,
606-8502, Japan,\\
$^d$ RIKEN BNL Research Center, BNL, Upton, NY 11973
}

\abst{
We show that the pseudogap 
of the quark density of states is formed in 
hot quark matter  as 
a precursory phenomenon of the color superconductivity
on the basis of  a low-energy effective theory.
We clarify that the soft mode of the di-quark pair field 
gives rise to a peculiar behavior of the quark dispersion relation
 and a short life-time of the quasiparticles
 near the Fermi surface, both of which make a depression of  the
density of states of quarks.
Our result suggests that the appearance of the pseudogap is 
a universal phenomenon  of strong coupling
superconductors, irrespective of the dimensionality.
}

\begin{document}

\maketitle

\section{Introduction}

It is one of the central issues in hadron physics
to determine the phase structure
of QCD at large  chemical potential $\mu$ and relatively 
low temperature $T$. 
The recent renewed interest 
in the color superconductivity (CS)\cite{ref:BL}
stimulated intensive studies in these region,
which in turn are revealing rich physics of the 
high density hadron/quark matter with CS\cite{ref:review}.

Possible physical realizations of the CS in compact stars or
ultrarelativistic heavy-ion collisions
are also discussed actively.
Here, note that these systems are at relatively low density $\rho$
where the strong coupling nature of QCD may show up.
The strong coupling may invalidate the mean-field approximation \`a la
BCS theory\cite{ref:BL},
and make the so-called Ginzburg region so wide
that precursory fluctuations of
the pair field can have a prominent strength and
may give rise to physically significant effects
even above the critical temperature $T_c$\cite{ref:KKKN}.

The existence of the large fluctuations 
suggests us that the CS may share some basic properties
with  the high-$T_c$
superconductivity (HTSC) of cuprates rather than with 
the usual superconductivity in metals.
One of the most characteristic phenomena of HTSC is the
existence of the {\em pseudogap}, i.e., 
the anomalous depression of 
density of state (DOS) $N(\omega)$ as a
function of the fermion energy $\omega$ around the Fermi surface
above $T_c$.
Although the mechanism of the pseudogap in HTSC is
still controversial,
precursory fluctuations of the pair field 
seem to be basic ingredients to realize the
pseudogap\cite{ref:HTSC}.
Thus, one may naturally expect that 
the pseudogap of the quark density of states
exists as a precursory phenomenon of the CS at finite $T$.
In this talk,  we shall show that  
it is the case using a chiral model\cite{ref:pgap}.

\section{Formalism}

To describe a system at relatively low $T$ and $\rho$,
it is appropriate to adopt a low-energy effective theory of QCD.
Here we employ the Nambu-Jona-Lasinio model
with the scalar-diquark interaction in the chiral limit,
\begin{eqnarray}
\label{eqn:lag}
{\cal L} &=& \bar{\psi}i/\hspace{-2mm}\partial \psi 
+ G_C \sum_A (\bar{\psi}i\gamma_5\tau_2 \lambda_A \psi^C)
(\bar{\psi}^C i\gamma_5 \tau_2 \lambda_A \psi) 
+ G_S[(\bar{\psi}\psi)^2+(\bar{\psi}i\gamma_5 \vec{\tau}\psi)^2],
\end{eqnarray}
where $\psi^C \equiv C\bar{\psi}^T$, 
with $C = i\gamma_2\gamma_0$.
 Here, $\tau_2$ and $\lambda_A$ mean 
the antisymmetric flavor SU(2) and color SU(3) matrices, respectively.
The coupling $G_S$ and the three dimensional momentum 
cutoff $\Lambda=650$ MeV are determined 
so as to reproduce the physical quantities and 
we choose $G_C= 3.11{\rm GeV}^{-2}$.

We neglect the gluon degrees of freedom, especially
their fluctuation, which is known to make the CS phase transition
first order in the weak coupling region\cite{ref:BL}.
However, nothing definite is known on the characteristics of the CS
in the intermediate density region.
In this work, simply assuming that the fluctuation of the
pair field dominates that of the gluon field,
we examine the effects of the precursory fluctuations of 
the diquark pair field on the quark sector in the T-matrix approximation 
(T-approximation).

The DOS $N(\omega)$ is given by
\begin{eqnarray}
N(\omega) &=& 4\int \frac{d^3 {\bfk}}{(2\pi)^3}
{\rm Tr}_{\rm c,f}\left[ \rho_0 ({\bfk},\omega) \right],
\label{eq:rho_0}
\end{eqnarray}
where $\rho_0 = (1/4){\rm Tr}[\gamma_0 \cal A]$ with
${\cal A}({\bfk},\omega)=-1/\pi \cdot{\rm Im}G^R ({\bfk},\omega)$
denoting the spectral function of a single quark.
The retarded Green function $G^R$ is given by the analytic continuation
of the imaginary-time Green function
${\cal G}$, which obeys the following Dyson-Schwinger equation
\begin{eqnarray}
{\cal G}({\bfk}, \omega_n)
= {\cal G}_0({\bfk}, \omega_n)\{1 + 
\tilde{\Sigma}({\bfk},\omega_n){\cal G}({\bfk},\omega_n)\},
\label{eqn:DS-Eq}
\end{eqnarray}
where $
{\cal G}_0 ({\bfk},\omega_n)
= [(i\omega_n+\mu)\gamma^0 - \bfk \cdot \bfgamma]^{-1}$
and$ \tilde{\Sigma}({\bfk},\omega_n)$ 
denote the free Green function and the self-energy in the 
imaginary time with $ \omega_n = (2n+1)\pi T $.

As was shown in \citen{ref:KKKN},
 the fluctuating diquark pair field
develops a collective mode
(the {\em soft mode} of the CS)
at $T$ above but in the vicinity of $T_c$, in accordance with the
Thouless criterion\cite{ref:Thoul}.
Our point in this work 
is  that the soft mode  in turn
contributes to  the self-energy of the quark field, thereby
can modify the DOS so much to give rise to a pseudogap.

The  quark self-energy $\tilde{\Sigma}$ owing to the soft mode
 may be  obtained by 
the infinite series of the ring diagrams shown in Fig.~1;
\begin{eqnarray}
\tilde{\Sigma}({\bfk},\omega_n) 
&=& T\sum_{n_1}\int \frac{d^3 {\bfk}_1}{(2\pi)^3}
\tilde\Xi ({\bfk}+{\bfk}_1, \omega_n+\omega_{n_1})
{\cal G}_0 ({\bfk}_1,\omega_{n_1}), \label{eqn:SE} \\
\tilde{\Xi} ({\bfk},\nu_n)
&=& -8G_C
\left( 1+G_C {\cal Q}({\bfk},\nu_n) \right)^{-1}, \label{eqn:PF}
\end{eqnarray}
with the lowest  particle-particle correlation function 
${\cal Q}({\bfk},\nu_n)$ \cite{ref:KKKN} and 
 $\nu_n = 2n\pi T$.

Inserting Eqs. (\ref{eqn:SE}) and (\ref{eqn:PF}) 
into Eq. (\ref{eqn:DS-Eq}) and performing the analytic continuation 
to the upper half of the complex energy plane, 
we obtain the retarded Green function,
$
G^R(\bfk,\omega) 
= ( G^{-1}_0(\bfk,\omega+i\eta) -\Sigma^R(\bfk,\omega) )^{-1},
$
with 
$\Sigma^R({\bfk},\omega) 
= \tilde{\Sigma}(\bfk,\omega_n)|_{i\omega_n=\omega + i\eta}$.
Here, the self-energy $\Sigma^R$ has the matrix structure
$\Sigma^R( \bfk ,\omega )
= \Sigma_0( \bfk ,\omega ) \gamma^0 
- \Sigma_{\rm v}( \bfk ,\omega ) \hat{\bfk} \cdot \bfgamma$
$\equiv \gamma^0(\Sigma_- \Lambda_- + \Sigma_+ \Lambda_+)$,
where $\Lambda_\mp = ( 1 \pm \gamma^0 \bfgamma\cdot\hat{\bfk})/2$
denotes the projection operators onto
the positive and negative energy states.
$\Sigma_{\mp}=\Sigma_0\mp\Sigma_{\rm v}$ represents
the self-energies of the particles and anti-particles,
respectively.
\begin{figure}[tb]
\begin{center}
\includegraphics[scale=.2]{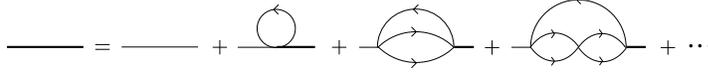}
\caption{
The Feynman diagrams representing the
quark Green function.
}
\label{fig:self}
\end{center} 
\end{figure}
\begin{figure}[b]
\begin{center}
\includegraphics[scale=.5]{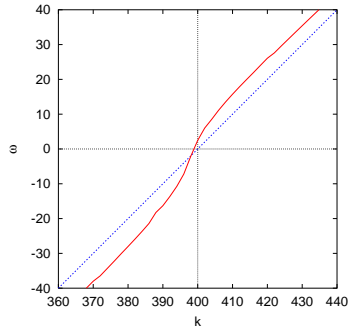} 
\caption{
The quark dispersion relation $\omega=\omega(k)$.
One observes a rapid increase of $\omega(k)$ around 
the Fermi momentum $k=400$MeV.
}
\label{fig:sig}
\end{center} 
\end{figure}

\begin{figure}[t]
\begin{center}
\begin{tabular}{cc}
\includegraphics[scale=0.55]{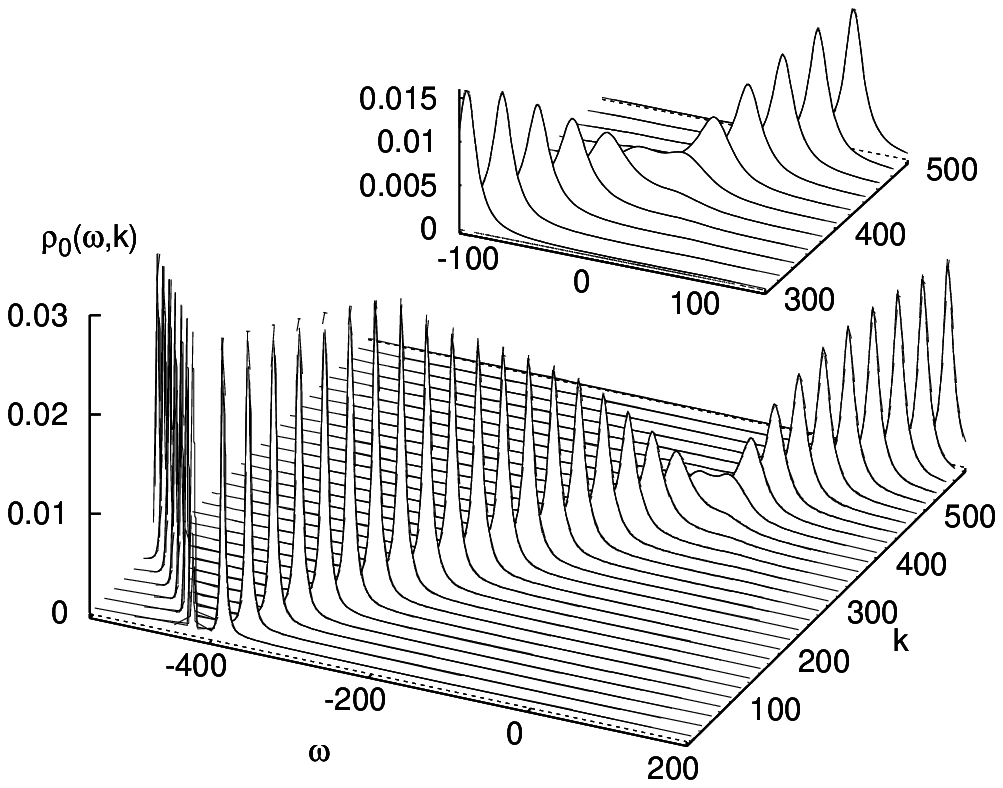} &
\includegraphics[scale=.85]{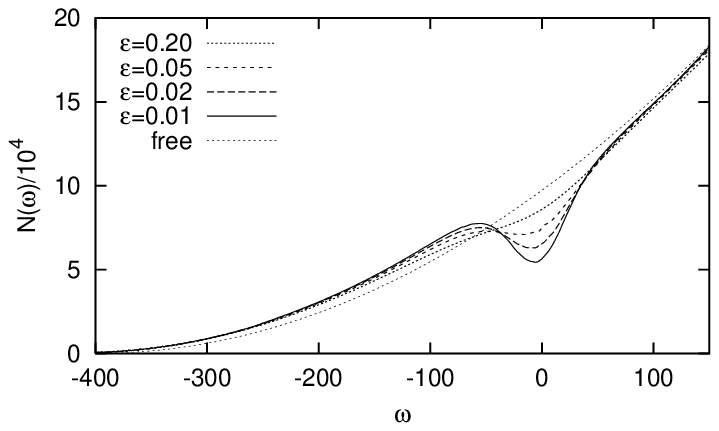} 
\end{tabular}
\caption{
{\bf Left panel:} 
The spectral function $\rho_0$.
There is a depression around $ \omega=0 $,
which is responsible for the pseudogap formation.
{\bf Right panel:} 
Density of state at $\mu=400$MeV and various $\varepsilon\equiv(T-T_C)/T_C$.
A clear pseudogap structure is seen, which survives
up to $ \varepsilon\approx 0.05$.
}
\label{fig:spc}
\end{center} 
\end{figure}

\section{Numerical Results and Discussions}

Since $\rho_0(\bfk, \omega)$ for $\omega >-\mu$
is well approximated solely by the positive-energy part,
we see the characteristic properties of 
the quark  self-energy $\Sigma_-$.
Numerical calculation shows that
${\rm Re} \Sigma_-$ shows a rapid increase
around the Fermi energy $ \omega=0$ at $ k=k_F $.
The behavior of ${\rm Re} \Sigma_-$ is responsible for 
the quark  dispersion relation $\omega=\omega(k)$.
We show a typical behavior of the quark dispersion relation
at $\mu=400$MeV and the reduced temperature
$ \varepsilon \equiv ( T-T_c )/T_c =0.01 $
in Fig. \ref{fig:sig}.
One sees a rapid increase of $\omega(k)$ around the Fermi momentum
$k=400$MeV, and hence $\partial \omega(k)/ \partial k$ becomes large
around this momentum.
Therefore, the density of states proportional
to $ (\partial \omega_-/\partial k)^{-1}$
becomes smaller near the Fermi surface, which suggests the
existence of a pseudogap.
Numerical calculation also shows that 
there is a peak of $|{\rm Im} \Sigma_-|$ around the Fermi energy,
which implies that the quasiparticles
around this energy are dumped modes.
${\rm Im} \Sigma_-$ describes 
a decay process of a quark to a hole and a diquark, 
q$\to$h$+$(qq), and this process is enhanced around $\omega=0$.

The spectral function
 $\rho_0(\bfk, \omega)$ is shown in
the left panel of Fig.~\ref{fig:spc},
at the same $\mu$ and $\varepsilon$ as those in Fig.~\ref{fig:sig}.
One can see the quasiparticle peaks of the quarks and anti-quarks
at $ \omega = \omega_-(k) \approx k-\mu $ and $ \omega = -k-\mu $,
respectively.
Notice that
the quasiparticle peak has a clear depression around $ \omega=0$.
The mechanism for the depression is easily understood
in terms of the characteristic properties of ${\rm Im} \Sigma_-$
mentioned above.

Integrating $\rho_0$, one obtains the DOS $N(\omega)$:
the right panel of Fig.~\ref{fig:spc} shows the DOS at $\mu=400$MeV
and various values of the reduced temperature $\varepsilon$
 together with that of the free quark system, $N_0(\omega)$.
As anticipated,
one can see a remarkable depression of $N(\omega)$,
i.e., the {\em pseudogap}, around 
the Fermi energy $ \omega=0 $;
$N(\omega)/N_0(\omega)|_{\omega=0} \simeq 0.55$
at $ \varepsilon=0.01 $.
The clear pseudogap structure
survives even at $ \varepsilon=0.05$.
One may thus conclude that
there is a  pseudogap region
within the QGP phase above $T_c$ up to
$T=(1.05\sim1.1)T_c$ at $\mu=400$MeV, for instance.
This wide range of $T$ may be just a reflection of the
strong coupling nature of the QCD at intermediate density region.
Our result obtained for a three-dimensional 
system tells us  that
a considerable pseudogap can be 
formed without the help of
the low-dimensionality as in the HTSC and
that the pseudogap phenomena in general
may be  universal  in any strong coupling
superconductivity.

In this work,
we have found that the pseudogap can be formed 
as a precursory phenomenon of the CS
in a rather wide region of $T$ above $T_c$.
It should be noted that
our work is the first calculation to show the formation
of the pseudogap in the relativistic framework.


%

\end{document}